\documentclass[pre,twocolumn,showpacs]{revtex4-1}
\usepackage{graphicx}

\newcommand{\vbar}{{\overline v}}
\newcommand{\Deff}{D_{\mathrm{eff}}}
\newcommand{\half}{\frac{1}{2}}

\newcommand{\myav}[1]{\langle{#1}\rangle}

\newcommand{\eqref}[1]{(\ref{#1})}
\newcommand{\Eqref}[1]{Eq.~\eqref{#1}}
\newcommand{\Eqsref}[1]{Eqs.~\eqref{#1}}
\newcommand{\Figref}[1]{Fig.~\ref{#1}}
\newcommand{\Figsref}[1]{Figs.~\ref{#1}}

\newcommand{\latin}[1]{{\itshape #1}}
\newcommand{\cf}{\latin{cf.}}

\newcommand{\via}{\latin{via}}
\newcommand{\viz}{\latin{viz.}}
\newcommand{\ie}{\latin{i.$\,$e.}}

\newcommand{\etal}{\latin{et al.}}

\newcommand{\adhoc}{\latin{ad hoc}}

\newcommand{\german}[1]{{\itshape #1}}
\newcommand{\ansatz}{\german{ansatz}}

\newlength{\figwidth}
\begin{document}

\title{Beading instability and spreading kinetics in
  grooves with convex curved sides}

\author{Patrick B. Warren}

\email{Email: patrick.warren@unilever.com}

\affiliation{Unilever R\&D Port Sunlight, Quarry Road East, Bebington,
  Wirral, CH63 3JW, UK.}

\date{December 30, 2015 }

\begin{abstract}
The coarsening kinetics for the beading instability for liquid
contained in a groove with convex curved sides (for example between a
pair of parallel touching cylinders) is considered as an open channel
flow problem.  In contrast to a V-shaped wedge or U-shaped
microchannel, it is argued that droplet coarsening takes place by
viscous hydrodynamic transport through a stable column of liquid that
coexists with the droplets in the groove at a slightly positive
Laplace pressure.  With some simplifying assumptions, this leads to a
$t^{1/7}$ growth law for the characteristic droplet size as a function
of time, and a $t^{-3/7}$ law for the decrease in the droplet line
density.  Some remarks are also made on the spreading kinetics of an
isolated drop deposited in such a groove.
\end{abstract}


\maketitle

\section{Introduction}
\begin{figure}[b]
\begin{center}
\includegraphics[clip=true,width=\figwidth]{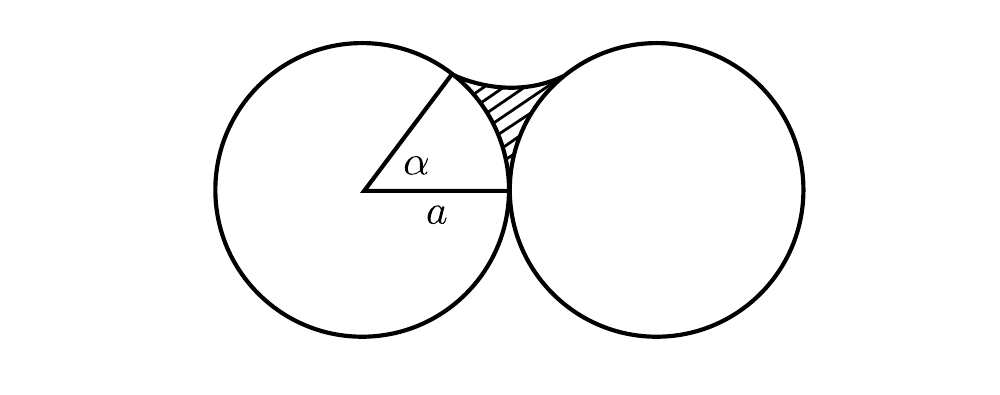}
\end{center}
\vskip -0.5cm
\caption{Liquid in a groove between parallel touching
  cylinders.\label{fig:geom}}
\end{figure}

Open channel flow problems have attracted much interest not only
because of possible applications in microfluidics \cite{SQ05, BDH+07,
  KHB+07, KBL+09, WBV10, BWV10, YKP+11, BSL+15, SBC+15}, but also
because of their relevance to liquids spreading on topographically
patterned surfaces such as human skin \cite{DAL03, CMR+09, SBH+11}.
The paradigmatic case of spreading in a V-shaped wedge has been
analysed both when liquid is supplied by a reservoir \cite{RY96,
  DAL03}, and in the starved (no reservoir) situation \cite{War04}.
Various aspects of these predictions have been confirmed
experimentally \cite{BDH+07, KHB+07, KBL+09, BSL+15}.  Flows in
U-shaped channels (\ie\ with concave sides) have also been considered
since such channels are easily micro-machined and are thus relevant
for microfluidics applications \cite{CMR+09, YKP+11}.  In the present
study, I revisit the problem, in the context of a groove with
\emph{convex} sides, such as that formed between a pair of parallel
touching cylinders (\Figref{fig:geom}).  In addition to possible
microfluidics applications, interest in this problem is motivated by
consideration of oily soil spreading along the fibres in textile yarns
in woven fabrics \cite{Pri70}.  It also has other potential
technological relevance, for example to molten solder wicking in
stranded copper wires and braids.

This problem throws up some interesting aspects not found in the
previous cases.  A groove with convex curved sides supports a stable
uniform liquid column at low loads, but at higher loading a uniform
liquid column may display a \emph{beading instability} in which it
breaks up into a string of droplets, similar to Rayleigh's observation
of the breakup of a thread of treacle on a paper surface \cite{Ray92}.
Although the statics are by now quite well understood \cite{Pri70,
  SC87, WBV10, BWV10, DBW+13, SBC+15}, I shall argue here that the
loading duality uniquely differentiates the present case from V-shaped
wedges and U-shaped microchannels, since the droplets (beads) can
coarsen by mass transport through connecting liquid columns.

Additionally, the loading duality implies an isolated droplet
deposited into such a groove may show several kinetic spreading stages
as it empties into the unfilled regions.  This aspect will be
discussed at the end.

\section{Wicking equation}
To start with, consider the general problem of open channel flow in a
channel of arbitrary cross section, and let $A(x,t)$ be the cross
section occupied by liquid.  A local mass conservation law holds
\cite{RY96, War04, RCR+08},
\begin{equation}
\frac{\partial A}{\partial t}+\frac{\partial(A\vbar)}{\partial x}=0\,.
\label{eq:conv}
\end{equation}
Herein, the mean flow rate $\vbar$ satisfies a Hagen-Poiseuille (HP)
law, $\vbar=-({k}/{\eta})\,{\partial p}/{\partial x}$, in which $k$ is
the permeability (a quantity with units of length squared,
\cf\ Darcy's law), $\eta$ is viscosity, and $p = p(A)$ is the
(loading-dependent) Laplace pressure.  Combining the HP law with
\Eqref{eq:conv} gives what can perhaps be called the \emph{wicking
  equation},
\begin{equation}
\frac{\partial A}{\partial t}=
\frac{\partial}{\partial x}\Bigl(
\frac{Ak}{\eta}\frac{dp}{dA}\,\frac{\partial A}{\partial x}\Bigr)\,.
\label{eq:wick}
\end{equation}
This is the basis for much of the subsequent development, and also
codifies the statics \via\ stability analysis.  It generally has the
character of a non-linear diffusion equation.  I have assumed that the
occupied cross section $A$ is weakly varying with $x$, so the
contribution to the Laplace pressure from the interface curvature in
the longitudinal direction can be neglected.  Note that the
permeability also depends on the loading, so that $k=k(A)$.

Insight can be gained by linearising about the uniformly loaded static
solution, \viz\ $A=A_0$ and $\vbar=0$.  Let us write
$A/A_0=1+\epsilon(x,t)$.  Then ${\partial \epsilon}/{\partial t}=
\Deff\,{\partial^2\epsilon}/{\partial x^2}$ where
$\Deff=({Ak}/{\eta})\,{dp}/{dA}$ is an effective diffusion
coefficient, evaluated at $A=A_0$.  If this is positive, then
perturbations will die away and the liquid will be self-levelling.  If
it is negative, then perturbations will grow indicative of the
aforementioned beading instability.  It is clear that the behavior
depends, not on the sign of Laplace pressure $p$, but rather on the
sign of $dp/dA$.  Physically, if $dp/dA>0$, an overfilled region will
have a higher Laplace pressure than an underfilled region, and the
liquid will flow to even things out.  On the other hand, if $dp/dA<0$,
liquid will flow from underfilled regions into overfilled regions,
magnifying the initial imbalance.  The interesting and unusual
property of a groove with convex curved sides is that both situations
occur, depending on the loading.

\section{Statics}
\begin{figure}
\begin{center}
\includegraphics[clip=true,width=\figwidth]{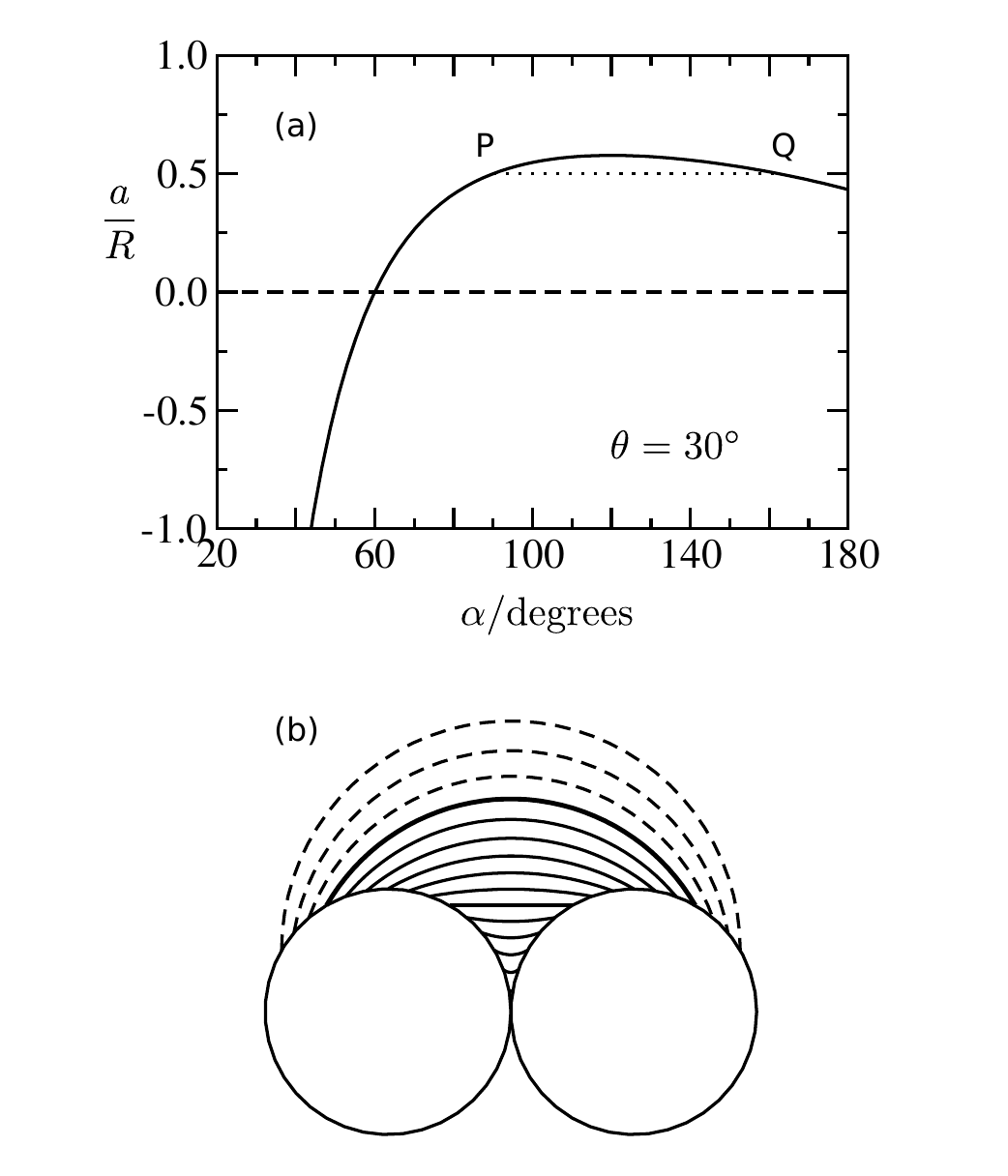}
\end{center}
\vskip -0.5cm
\caption{(a) The dimensionless Laplace pressure as a
  function of the filling angle, with liquid-solid contact angle
  $\theta=30^\circ$.  (b) Filling stages in this problem: the
  thick solid line is the maximum Laplace pressure at
  $\alpha=120^\circ$; uniform loading above this (dashed lines) is
  predicted to be unstable. \label{fig:lap}}
\end{figure}

A specific example of the loading duality is provided by the groove
between a pair of parallel touching cylinders, shown in
\Figsref{fig:geom} and~\ref{fig:lap}.  Since I neglect the interface
curvature in the longitudinal direction, the transverse profile of the
free surface is characterised by an arc of a circle with radius $R$.
Taking $R>0$ to indicate the surface is convex outwards, one has
$p=\gamma/R$ where $\gamma$ is surface tension.  In this problem it is
convenient \cite{SBC+15} to parametrise the loading by the wrapping
angle $\alpha$, shown in \Figref{fig:geom}. Note that the area is a
monotonically increasing function of $\alpha$, so $dA/d\alpha>0$.

Elementary trigonometric arguments, first presented to my knowledge by
Princen \cite{Pri70}, show that
\begin{equation}
\frac{a}{R}=-\frac{\cos(\theta+\alpha)}{1-\cos\alpha}\label{eq:abyR}
\end{equation}
where $\theta$ is the liquid-solid contact angle.  This result holds
for both convex and concave interfaces, and the sign has been inserted
in accord with the above convention that $R$ is positive if the
interface is convex.

From this we find (after
a little rearrangement)
\begin{equation}
\frac{dR}{d\alpha}=-\frac{R^2\cos(\half\alpha+\theta)}
{2a\sin^3(\half\alpha)}\,.\label{eq:drda}
\end{equation} 
Since $p\propto a/R$, \Eqsref{eq:abyR} and \eqref{eq:drda} show that
$p$ increases through zero at $\alpha+\theta=\half\pi$ to reach
a weak maximum at $\half\alpha+\theta=\half\pi$
(\ie\ $\alpha=\pi-2\theta$).  A specific example is shown in
Fig.~\ref{fig:lap}a, for a contact angle $\theta=30^\circ$.  The zero
crossing is at $\alpha=60^\circ$ and the Laplace pressure maximum is
at $\alpha=120^\circ$.  Fig.~\ref{fig:lap}b shows a selection of the
corresponding filling states.  The stability requirement that
$dp/dA\propto dp/d\alpha>0$ indicates that filling states with
$\alpha>120^\circ$ are unstable with respect to the above-mentioned
beading instability \cite{heightnote}.

\section{Kinetics}
\subsection{Beading instability}
As we have seen for the case of parallel touching cylinders, the
Laplace pressure in a groove with convex sides may show a maximum as a
function of the loading.  Beyond the maximum, $dp/dA<0$, and therefore
a beading instability arises in which the uniformly loaded state is
unstable towards the growth of perturbations.  But first, what could
be the final state in such a situation?  In principle one can have two
different states of loading at same Laplace pressure, which can
therefore be in coexistence, for example points P and Q in in
Fig.~\ref{fig:lap}b.  However the higher loaded state is always in the
unstable region.  The logical conclusion is that the excess liquid is
expelled into a large droplet that sits somewhere on the two
cylinders, coexisting with a stable column of liquid in the groove.
In the case of parallel touching cylinders, the wrapping angle for
this stable liquid column would have to satisfy
$\half\pi-\theta<\alpha<\pi-2\theta$.  The first inequality arises
because the liquid column coexists with a large drop at a (weak)
positive Laplace pressure \cite{dimnote}.  The second inequality is
the column stability condition.

I turn now to the coarsening kinetics.  A uniformly overloaded state
will certainly break up into a string of droplets, but in a groove
with convex curved sides these droplets are always connected by liquid
columns with a finite filling depth according to the above argument.
Therefore the larger droplets can eat the smaller ones, by
transporting liquid along the connecting liquid columns.  This stands
in contrast to the V-shaped wedge for example, where spatially
separated droplets are disconnected \cite{CF69, RDN99, KBL+09} and the
droplet population has to coarsen by some other mechanism, such as a
prewetting film \cite{PRW00, RP07, MP13}, or transport through the
vapor phase in the case of a volatile liquid.

With a simplifying assumption about how the Laplace pressure depends
on the droplet size, aspects of the Lifshitz-Slyozov-Wagner (LSW)
theory of droplet coarsening \cite{Bra94} can be adapted to the
present case.  The simplifying assumption is that the Laplace pressure
scales inversely with droplet size as $p\propto V^{-1/3}$, where
$V\equiv R^3$ is the droplet volume.  This is likely to be true only
asymptotically \cite{Car76, dGB04}.  Nonetheless, let us suppose that
this is true, and the situation has evolved so that there is a string
of droplets sitting in the groove.  The following mean-field scaling
\ansatz\ predicts how the characteristic droplet size $R$, and
characteristic spacing between droplets $L$, evolve with time.

\begin{figure}
\begin{center}
\includegraphics[clip=true,width=\figwidth]{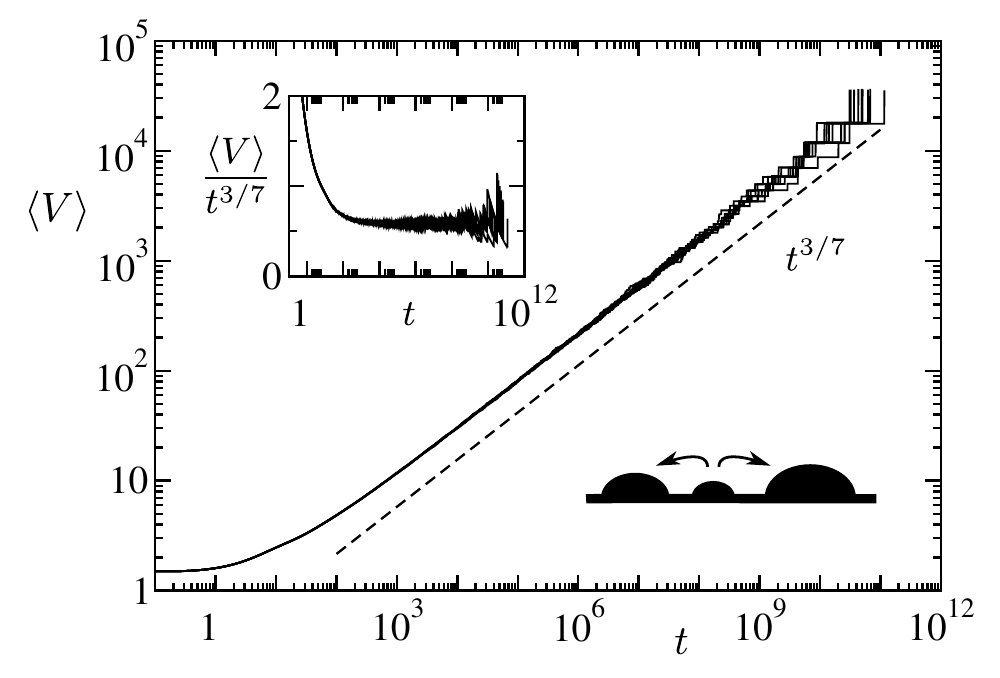}
\end{center}
\vskip -0.5cm
\caption{Mean droplet volume as a function of time.  Droplets coarsen
  by viscous hydrodynamic transport along the connecting liquid columns
  (diagrammatic inset), according to the rules prescribed in the
  Appendix.  Results from 10 independent simulation runs are shown.
  Each simulation was initialised with $5\times 10^4$ equispaced
  droplets, with random sizes taken from a uniform distribution
  $V_0<V<2V_0$.  Droplets are removed when $V<0.9\,V_0$, and their
  liberated contents are added to the liquid column ($\beta=1$ in the
  model).  The time step parameter was
  $\epsilon=0.02$.  \label{fig:coarsen}}
\end{figure}

First, the mass flux between adjacent droplets will be
\begin{equation}
  J\sim\frac{a^4}{\eta}\,\frac{\Delta P}{L}\,.\label{eq:s1}
\end{equation}
This just expresses the HP law in scaling form.  Shown here is the
prefactor for the case of parallel touching cylinders, where the
fourth power of the cylinder radius $a$ arises from the product of the
Darcy permeability $k\sim a^2$ and the cross sectional area of the
connecting liquid column $A\sim a^2$.  The analysis does not depend on
this particular geometry though.

The mass flux is incorporated into a \emph{local} mass conservation law,
\cf\ \Eqref{eq:conv}, which governs the growth of the mean droplet volume,
\begin{equation}
  \frac{dV}{dt}\sim J\,.\label{eq:s2}
\end{equation}
At the same time there is a \emph{global} mass conservation law which
relates the mean droplet size to the mean spacing,
\begin{equation}
  \frac{V}{L}\sim\omega\sim\mathrm{const}\,,\label{eq:s3}
\end{equation}
where $\omega$ is the mass per unit length \cite{initnote}.

Combining \Eqsref{eq:s1}--\eqref{eq:s3}, together with $V\equiv R^3$
and $\Delta p\sim \gamma/R$, shows that
\begin{equation}
R^2\,\frac{dR}{dt}\sim \frac{\gamma a^4\omega}{\eta R^4}\,.
\end{equation}
This integrates to
\begin{equation}
  R\sim\Bigl(\frac{\gamma a^3\omega t}{\eta}\Bigr)^{1/7}\,.
\end{equation}
Thus the prediction is that the mean droplet size should grow as
$t^{1/7}$ and the droplet line density (\ie\ $1/L\sim R^{-3}$) should
diminish as $t^{-3/7}$.

\begin{figure}
\begin{center}
\includegraphics[clip=true,width=\figwidth]{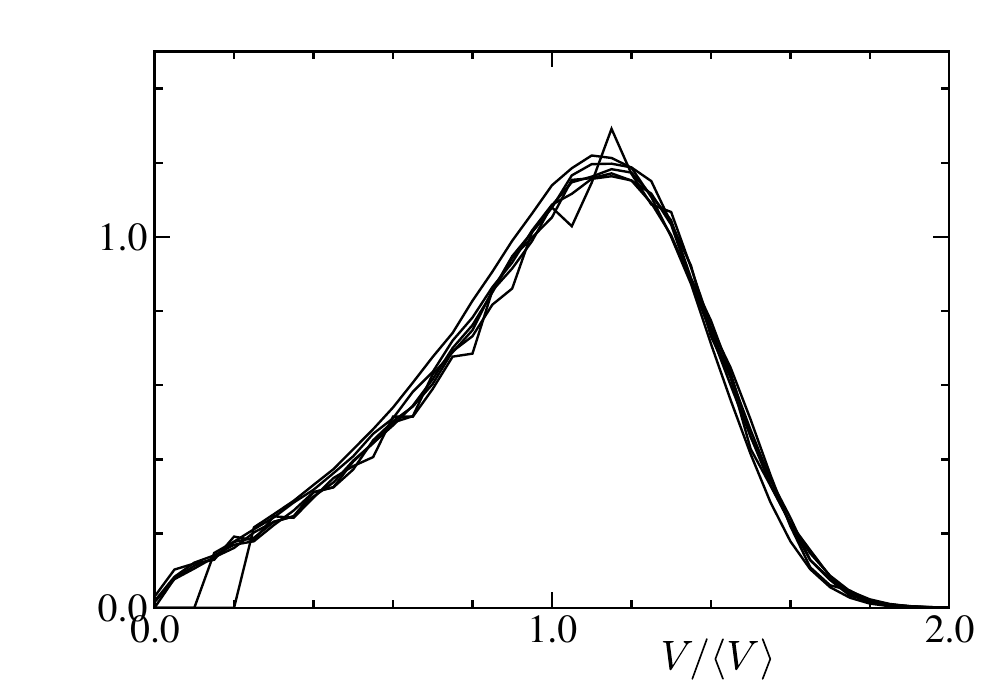}
\end{center}
\vskip -0.5cm
\caption{Scale invariance of droplet size distribution.  Histograms
  are computed when there are 10\,000, 5000, 1000, 500, 300 and 100
  droplets remaining out of an initial $5\times 10^4$, combining
  data from 100 independent simulation runs.  Simulation parameters as
  for \Figref{fig:coarsen}.\label{fig:dropsize}}
\end{figure}

\begin{figure}
\begin{center}
\includegraphics[clip=true,width=\figwidth]{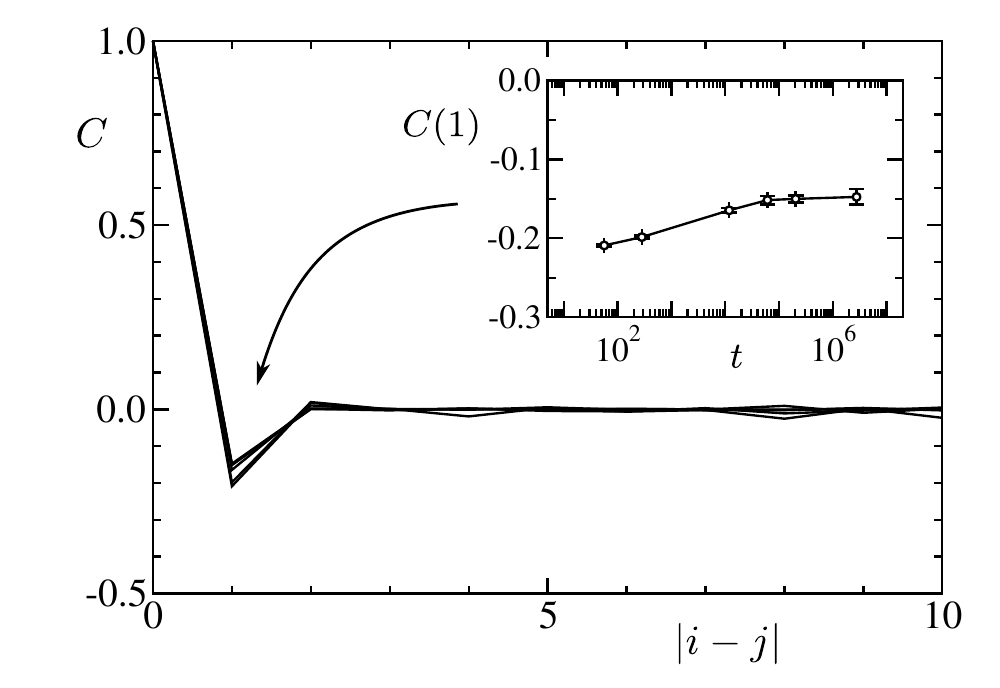}
\end{center}
\vskip -0.5cm
\caption{Equal-time size correlation function, $C(|i-j|)$, computed
  for the same set of simulations used for \Figref{fig:dropsize}.  The
  inset shows the time dependence of the depth of the
  nearest-neighbour minimum.  Error bars are from block averaging (10
  blocks $\times$ 10 runs).\label{fig:corrn}}
\end{figure}

The mean-field assumption is questionable given the one-dimensional
nature of the coarsening problem.  To investigate specifically just
this aspect, I undertook numerical simulations using the model
described in the Appendix.  This confirms that the mean-field scaling
\ansatz\ does indeed predict how the mean droplet size grows with time
(\Figref{fig:coarsen}), and also demonstrates LSW-like scale
invariance for the droplet size distribution (\Figref{fig:dropsize}).

The origin of the mean-field behaviour is apparent if one examines the
equal-time droplet size correlation function, $C(|i-j|)$, shown in
\Figref{fig:corrn}. As can be seen the only significant correlation
appears to be between nearest neighbours, at $|i-j|=1$.  This
indicates that droplets which are larger than average tend to be
adjacent to droplets which are smaller than average, but apart from
this no significant long-range correlations develop.

To see the origin of this nearest-neighbour correlation hole, consider
an artificial situation in which one large droplet sits in a uniform
string of equisized smaller droplets.  Away from the large droplet,
the ambient Laplace pressure is uniform and no coarsening takes place.
However the large droplet has a sub-ambient Laplace pressure and so
starts to draw in material from its immediate neighbours.  This causes
the immediate neighbours to shrink, and increases their Laplace
pressure relative to the ambient background.  The next-nearest
neighbours then see that these droplets have started to shrink, and so
they in turn start to grow.  The process continues, and it is easy to
see that it will generate a staggered array of droplet sizes, with a
concomitant nearest-neighbour negative size correlation.

The nearest-neighbour correlation hole diminishes somewhat as time
progresses, but eventually appears to settle down to a value
$C(1)\approx -0.15$ (\Figref{fig:corrn} inset).  One should point out
that the coarsening dynamics in the model is quite subtle, for
instance, the largest droplet at time $t$ may not necessarily be the
largest droplet at some later time $t'>t$.  This is because the growth
rate of a droplet depends not only on its size but also on the sizes
of its neighbours, and how far away they are.

\subsection{Droplet spreading}
The remaining point of kinetic interest concerns the fate of a droplet
of liquid deposited onto an initially empty groove. The droplet will
start to empty into the groove, presumably driving a
Bell-Cameron-Lucas-Washburn (BLCS) type flow from what is in effect a
shrinking droplet reservoir \cite{RY96, DAL03, RCR+08}.  This should
persist all the way until the loading falls below the Laplace pressure
maximum.  Past this point, by analogy to the V-shaped wedge
\cite{War04}, one expects the spreading rate to slow down since the
reservoir has been exhausted.

For the case of parallel touching cylinders, a second power law
appears at a very late stage where everywhere $\alpha\ll1$.  In this
limit, the wetted portion of the groove has shrunk to a narrow fissure
with a width of the order $R\sim\alpha^2a$ (see \Eqref{eq:abyR} in the
limit $\alpha\to0$) and a depth of the order $\alpha a$.  One
therefore expects $A\sim \alpha^3a^2$, and presumably
$k\sim\alpha^4a^2$ since the permeability should largely be determined
by the width.  Similar to the coarsening kinetics problem, a scaling
analogue of the HP law can be introduced.  In the present case this is
${dL}/{dt}\sim ({k}/{\eta})\times{\Delta p}/{L}$ where $L$ is the
length of the wetted portion of the groove and $\Delta p\sim \gamma/R$
is the Laplace pressure.  Substituting the above scaling expressions
gives ${dL}/{dt}\sim {\gamma\alpha^2 a}/({\eta L})$.  An additional
constraint comes from the analogue of the global mass conservation law
in \Eqref{eq:s3}, namely that the total droplet volume $\Omega
\sim\alpha^3a^2L$ should be conserved. Eliminating $\alpha$ between
this volume constraint and the HP scaling law yields ${dL}/{dt}\sim
{\gamma\Omega^{2/3}}/({\eta a^{1/3}L^{5/3}})$.  This integrates to the
final rather esoteric result
\begin{equation}
L\sim({\gamma
  t}/{\eta})^{3/8}\,\Omega^{1/4}\,a^{-1/8}\,.
\end{equation}
In other words, the initial $L\sim t^{1/2}$ spreading law (BLCS)
should weaken when the droplet reservoir vanishes, and eventually
enter an $L\sim t^{3/8}$ power law in the final starved state.

\section{Discussion}
I have argued that the kinetic aspects of spreading and (de)wetting
for a liquid contained in a groove with convex curved sides presents
some unique aspects when compared, for example, to a V-shaped wedge or
a U-shaped microchannel.  The novel aspects arise from an underlying
loading duality, wherein a liquid column is stable at low loading, but
becomes unstable at higher loading.

For the case of uniform loading above the critical loading threshold,
a beading instability should be observed in which the liquid column
breaks up into a string of droplets, which subsequently coarsen by
mass transport along connecting liquid columns.  With a simplifying
assumption about the dependence of the Laplace pressure on the droplet
volume, a mean-field scaling \ansatz\ indicates that the droplet size
and line density scale with non-trivial power laws in time, and
simulations show that this is not destroyed by the one-dimensional
nature of the problem.

Of course, the simplifying assumption about the Laplace pressure
scaling does not hold in reality, and in general I would expect the
clean power law behaviour to be modified by finite-droplet-size
effects, which may be quite persistent.  Some general predictions
should be robust however, such as the relatively slow droplet growth
\via\ transport along connecting liquid columns, and the negative
correlation between nearest neighbour droplet sizes shown in
\Figref{fig:corrn}.  These could perhaps be tested in an
electrowetting experiment \cite{KHB+07, KBL+09}.  I should caution
that the predicted slow coarsening kinetics may be overtaken by other,
ultimately faster, mechanisms.  For example transport through the
vapour phase for a volatile liquid may ultimately lead to a $R\sim
t^{1/3}$ coarsening law, as in LSW theory \cite{Bra94}.

Another prediction arising from the loading duality is that a droplet
deposited in the groove should show a staged spreading kinetics,
starting with the classic Bell-Cameron-Lucas-Washburn law as the
droplet initially acts like a reservoir, slowing when the loading
falls everywhere below the critical value, and possibly ending with a
new power law in the final starved state.  Again, this may perhaps be
probed experimentally.
 
\appendix
\section{Simulation of coarsening kinetics}\label{app:sim}
I introduce a simplified model of the coarsening kinetics to test
specifically the mean-field \ansatz\ presented in the main text.  In
the model, I consider a one-dimensional string of $i=1\dots N$ droplets,
connected by liquid columns as indicated in the lower inset in
\Figref{fig:coarsen}.  Big droplets grow, and small droplets shrink,
under the influence of viscous hydrodynamic transport through the
liquid columns.  This is driven by differences in the Laplace pressure
between neighbouring droplets.  To establish a system of kinetic
equations for the droplet sizes, I suppose that the Laplace pressure
in the $i$-th droplet is proportional to $V_i^{-1/3}$, where $V_i$ is
the droplet volume (this is the simplifying assumption mentioned in
the main text).  According to the HP law, the Laplace pressure
difference drives a mass flux through the connecting liquid column as,
\cf\ \Eqref{eq:s1},
\begin{equation}
J_i=\frac{V_i^{-1/3}-V_{i+1}^{-1/3}}{L_i}\,.
\label{eq:a1}
\end{equation}
In this $L_i$ is the distance between the $i$-th and $(i+1)$-th
droplets, and all other material properties in the problem have been
subsumed into the definitions of length, volume and time.  Note that
$V_{i+1}>V_i$ implies $J_i>0$, so that liquid flows from smaller
droplets to larger droplets. Given the fluxes, mass conservation
dictates that, \cf\ \Eqref{eq:s2},
\begin{equation}
\frac{d V_i}{dt} = J_{i-1}-J_i\,.
\label{eq:a2}
\end{equation}

\Eqsref{eq:a1} and \eqref{eq:a2} are the required set of non-linear
kinetic equations.  Since they predict that droplets shrink, as well
as grow, we need a rule which governs how shrinking droplets can
disappear.  At this point it is convenient to introduce a fiducial
volume $V_0\sim R_0^3$, where the fiducial length $R_0$ is set by the
height of the connecting liquid column (in the simulations,
$V_0=R_0=1$). A simple rule for shrinking droplets is that they vanish
when $V_i<\alpha V_0$.  If this happens, the droplet is removed and the
distance between the remaining droplets is set equal to
$L_{i-1}+L_i+\beta V_i^{1/3}$, where the third term is an
\adhoc\ correction for the length contributed by the vanished droplet
(taking $\beta$ as a free parameter).

As an initial condition I set $V_i=r_i V_0$ where $r_i$ is a random
number chosen from a uniform distribution, $1\le r_i< r_m$.  The
droplets are initially equispaced, with $L_i=R_0$.  Periodic boundary
conditions are imposed.

The droplet volumes are evolved according to \Eqsref{eq:a1} and
\eqref{eq:a2}, alongside the above rule for removing droplets which
become too small.  \Eqsref{eq:a1} and \eqref{eq:a2} are integrated
using a simple, adaptive, Euler-type forward finite difference scheme,
with a time step $\Delta t$ chosen such $dV_i/dt \times \Delta t/V_i
\le \epsilon$, in other words so that the fractional change in any
droplet size does not exceed $\epsilon$ in any time step.  For the
reported simulations I used $\alpha=0.9$ and $\beta=1$ for the
vanishing rule, $r_m=2$ for the maximum initial drop size relative to
$V_0$, and $\epsilon=0.02$ for the choice of time step.  I have
checked the results are insensitive to these choices.

As time evolves, the larger droplets grow at the expense of the
smaller droplets, and the smallest droplets shrink and vanish.
Eventually the simulation stops when there is one large final droplet
($N=1$).  I monitor the mean droplet volume
$\myav{V}=(1/N)\sum_{i=1}^N V_i$ as a function of time (note that $N$
changes as droplets disappear), and at selected time points record the
drop size distribution.  Typical results, aggregated from multiple
simulation runs from independent starting points, are summarised in
\Figsref{fig:coarsen} and~\ref{fig:dropsize}.

I also calculate periodically the equal-time correlation function
\begin{equation}
  C(|i-j|)=\frac{\myav{\Delta V_i\,\Delta V_j}}%
  {\myav{\Delta V^2}}\,,\label{eq:cij}
\end{equation}
where $\Delta V_i=V_i-\myav{V}$ is the deviation from the mean droplet
volume, at time $t$.  This function is shown in \Figref{fig:corrn},
evaluated at various points in the simulation.


\begin{thebibliography}{30}
\bibitem [{ini()}]{initnote}%
  \BibitemOpen
  \href@noop {} {}\bibinfo {note} {It seems likely that $\omega\equiv R^3/L\sim
  a^2$ since this is the only length scale. This would correspond to an initial
  droplet volume of order $a^3$ and droplet spacing of order $a$.}\BibitemShut
  {Stop}%
\end{thebibliography}
%

\end{document}